          [2004/02/17]
\documentclass[a4paper,twocolumn]{esapub} % European paper

\usepackage{amssymb}
\usepackage{graphicx} 
\usepackage{pstricks}
\newrgbcolor{DarkBlue}{0.1 0.1 0.5}
\newrgbcolor{Red}{0.9 0.0 0.1}
\newrgbcolor{Green}{0.0 0.9 0.1}
\newrgbcolor{Gray}{0.5 0.5 0.5}
\newrgbcolor{Blue}{0 0 0.9}
\newrgbcolor{Black}{0 0 0}
\newrgbcolor{Orange}{0.95 0.94 0.80}

\usepackage{natbib}
\title{Sources of Cosmic Rays and Galactic Diffuse Gamma Radiation}
\author{Sabrina Casanova}
\affil{MPI f\"ur Radioastronomie, Auf dem H\"ugel 69, 53121 Bonn, Germany, casanova@mpifr-bonn.mpg.de}
\author{Peter L. Biermann}
\affil{MPI f\"ur Radioastronomie, Auf dem H\"ugel 69, 53121 Bonn, Germany}
\affil{Department of Physics and Astronomy, University of Bonn, Germany, plbiermann@mpifr-bonn.mpg.de}
\author{Ralph Engel}
\affil{Institut of Nuclear Physics, Forschungszentrum Karlsruhe, 76021 Karlsruhe, Germany, engel@ik.fzk.de}
\author{Athina Meli}
\affil{MPI f\"ur Radioastronomie, Auf dem H\"ugel 69, 53121 Bonn, Germany, ameli@mpifr-bonn.mpg.de}
\author{Ralf Ulrich}
\affil{MPI f\"ur Radioastronomie, Auf dem H\"ugel 69, 53121 Bonn, Germany, rulrich@mpifr-bonn.mpg.de}

\begin{document}


\maketitle

%%\begin{picture}(1,1)(-00,400)
%%\includegraphics[totalheight=12cm]{MPIfR-Logo.eps}
%%\end{picture}
%%\begin{picture}(1,1)(-1940,400)
%%\includegraphics[totalheight=12cm]{mpglogo.eps}
%%\end{picture}
%%\vspace{2cm}

%\vspace{2.5cm}

\begin{abstract} 
The diffuse galactic gamma-ray spectrum measured by the 
EGRET experiment \citep{Hunter:1997}
are interpreted within a scenario in which
cosmic rays (CRs) are injected by three different kind of sources,
(i) supernovae (SN) which explode into the interstellar medium (ISM),
(ii) Red Supergiants (RSG),
and (iii) Wolf-Rayet stars (WR), where the two latter explode into their 
pre-SN winds \citep{Biermann:2001iu,Sina:2001}. 
\end{abstract}

\section{Introduction}

\begin{table}[b]
\label{tab_sources}
\centering
\caption{Interaction spectra for the different type of supernovae.
$E_{\rm knee}$
and $E_{\rm cut-off}$ are in GeV. $\gamma_1$ and $\gamma_2$ are the spectral 
indices ($\Phi=\Phi_0 E^\gamma$) below and above 
$E_{\rm knee}$. $Z$ is the charge of the nucleus.}
\vspace{0.3cm}
\begin{tabular}{ccccc}
\hline
SN type&  $E_{\rm knee}$ & $E_{\rm cut-off}$ &$\gamma_1$& $\gamma_2$    \\
\hline
ISM       &                 & $3\,Z \,10^{5}$   & -2.75 &                   \\
RSG       &                 & $3\,Z \, 10^{5}$  & -2.33 &                    \\
WR        &  $2\,Z \,10^{6}$& $Z \, 10^{8}$     & -2.88 &      -3.21 \\
\hline
\end{tabular}
\end{table}
Conventional models of diffuse galactic gamma-ray production 
are based on three main processes.
Gamma-rays are 
produced through the decay of $\pi^0$ as secondary particles of 
hadronic collisions of CRs with the ISM, 
such as proton-hydrogen or proton-helium collisions,
in bremsstrahlung processes of CR electrons 
with the ISM, and through inverse Compton scattering of CR electrons
 with interstellar radiation fields. These models can explain
a wide range of observations like the energy spectra
 below $\approx 1$~GeV or the integrated flux from 
the outer parts of our Galaxy. Nonetheless 
the diffuse galactic gamma-rays observed by the EGRET experiment from the inner Galaxy
 above $\approx$~1~GeV exceeds by about 60$\,$\% 
the intensity predicted by these calculations (the measured spectrum is 
too hard).

In this contribution we shall present first results on the diffuse
gamma-ray production expected in the model by \citet{Biermann:2001iu}.
In this model, in addition to the aforementioned
processes, interactions of hadrons of the relatively 
hard CR injection spectrum (see Tab.~\ref{tab_sources}) 
with the pre-SN winds are considered as further possible sources of
diffuse galactic gamma-rays \citep{Biermann:2001iu}.

%In section I we present an overview of the method used to compute the
%diffuse galactic gamma ray radiation and discuss the nature of
%the various gamma producing processes. Cosmic ray spectra being an
%important input for the computation of gamma rays, we present 
\begin{figure}[t]
\centering
\includegraphics[width=7.5cm]{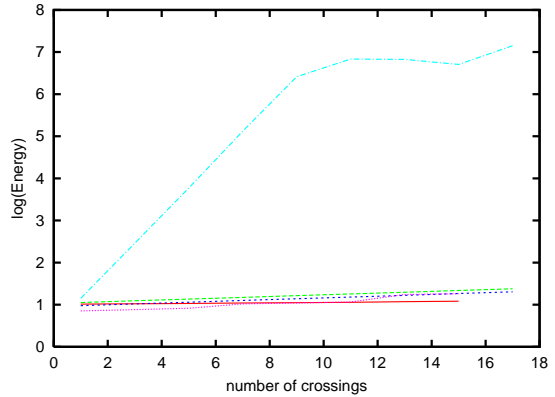}
\caption{The log of the mean energy of the CR particles versus the number of 
shock crossings for $V_{sh}=0.01c$. 
Starting from the bottom for $5^{o}, 25^{o}, 65^{o}, 80^{o}$ and $85^{o}$
respectively. We see the difference in the energy gain of CR particles for the
almost perpendicular case compared to smaller shock inclinations.
An effect that \citep{Jokipii:1987} pointed out as well.}
\label{fig_athina1}
\end{figure}

In section~2 a Monte Carlo code is developed to model particle acceleration in
non-relativistic near parallel and highly oblique shock configurations,
including cross-field diffusion, with application to RSG and WR winds.
In section~3 we review the adopted model of diffuse galactic gamma ray 
production and introduce all necessary input parameters. 
%the galactic distribution of the
%various species, the spectral and three dimensional distributions of cosmic rays producing
%gamma rays, and we introduce the cross sections for the various
%relevant scattering processes. 
In section~4 and 5, we present our results, compared to EGRET, CASA-MIA and
KASCADE data. 
%comprising the spectrum of gamma rays predicted and also the gamma
%ray distribution predicted as a function of longitude, and
%we compare these to existing observations. 
In section~6, we summarize
our findings and discuss some further improvements in our models.

\section{Sources of Cosmic Rays}
\citet{Jokipii:1987} investigated the rate of the energy gain and the maximum energy
in non-relativistic shocks, which can be attained in given conditions, such as the effect of a highly 
oblique magnetic field to the scattering of the CR particles. Briefly, he showed that 
if the perpendicular diffusion ($k_{\perp}$) is much smaller than the parallel one ($k_{\|}$), 
CR particles can gain considerable energy in quasi-perpendicular shocks compared to quasi-parallel ones. \\
We have constructed a Monte Carlo code in order to simulate non-relativistic near parallel
and oblique shocks with $0^\circ \lesssim \psi \lesssim 90^\circ$ where
$\psi$ is the angle between the shock normal and the magnetic field, seen in the shock frame.
The velocities of the upstream plasma flow are kept between $0.001c$ and $0.01c$, which correspond 
to astrophysical environments with non-relativistic shocks such as in WR winds, RSG winds, etc. 
In order to investigate the behavior of the particle scattering, between 
the quasi-parallel and oblique shock configurations, a cross-field diffusion is also allowed. 
Briefly, our Monte Carlo simulations in comparison to the Jokipii's theoretical work show the 
same trend (see Fig.~\ref{fig_athina1}). 
Furthermore, for realistic environments, such as 
RSG and WR stars, our calculations suggest that WR winds could
provide large enough radii for protons to reach enough energy in the available time and space, 
and we claim that the maximum energy (for Z=1) reached could be $\sim$ $10^{17}$ eV, in about   
$10^{10}$ sec ($\approx 10^4$yrs), which corresponds to $10^{19}$cm ($\approx$ 10pc).
On the other hand in RSG winds we may find the conditions (smaller radii and corresponding time) where the 
maximum energy could be about $3 \times 10^{14}$ eV. There is further work under way which 
calculates the real time versus energy, attainable in highly oblique non-relativistic shocks found in WR 
and RSG stars and investigates in detail the constraints put in these models, following 'limits' that 
may apply (e.g. $\kappa\; > \; r_g \; V_{sh}$) \citep{Meli:2004}.

\section{Models for Gamma-Ray Production}
\begin{figure}[t]
\centering
\includegraphics[width=7.5cm]{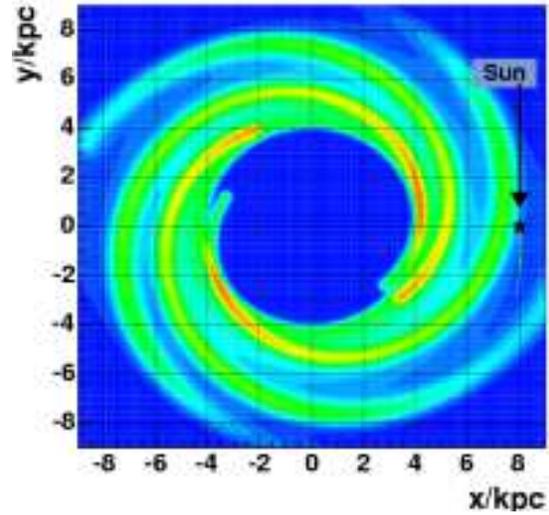}
\caption{Model of the ISM density based on COBE/DIRBE 
data \citep{Drimmel:2001}. The
logarithmic spiral structure is superimposed on a purely radial distribution
of molecular and ionized hydrogen \citep{Launhardt:2002tx, Kalberla:1998}.}
\label{fig_spirals}
\end{figure}[b]
\begin{figure}
\centering
\includegraphics[width=7.5cm]{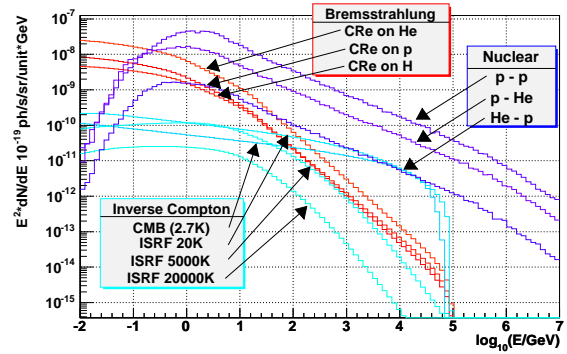}
\caption{Gamma-ray production spectra, see Eq.~(\ref{equ_IE}). The spectra are shown 
in units of photons per \textit{atom}, or
$cm^3$ in the case of inverse Compton.}
\label{fig_production}
\end{figure}
\begin{figure}[t]
\centering
\includegraphics[width=8.5cm]{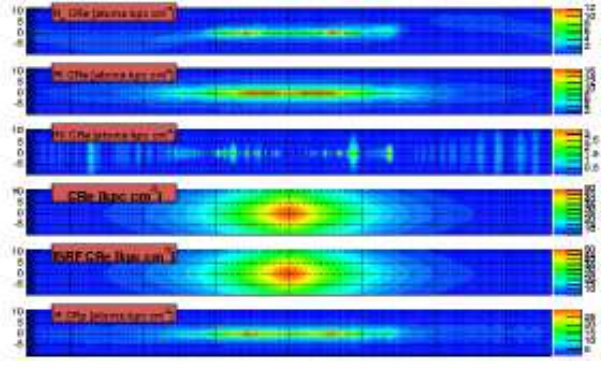}
\caption{Here the solution of $I_L(l,b)$ (Eq.~\ref{equ_IL}) is shown for
different combinations of CR and target particle types $k$,$T$.}
\label{fig_enhancement}
\end{figure}
The gamma-ray spectrum is computed using a three-dimensional spatial 
distribution of matter and 
radiation fields in our Galaxy. In this model the ISM is composed of
atomic ($\approx 1 \times 10^{9}M_{\odot}$), molecular 
($\approx 3\times 10^{9}M_{\odot}$) and ionized hydrogen
 ($\approx 10^{8}M_{\odot}$) as well as 10$\,$\%~He. 
Its spatial distribution is taken from
 \citet{Launhardt:2002tx, Kalberla:1998}. A logarithmic spiral arm 
model based on
COBE/DIRBE data is also incorporated (see Fig.~\ref{fig_spirals}) \citep{Drimmel:2001}. 
The interstellar photon densities of interest have been approximated by 
three diluted blackbody spectra from the interstellar radiation field (ISRF) 
plus the 2.7K microwave background (CMB) \citep{Bloemen:1985}.\\
CR spectra and their spatial distributions are another input for the 
calculations. We assume CR protons and electrons to be radially 
distributed with a radial exponential scale $a=5$ kpc. The assumed
exponential scale heights $h_{CR}$ are 0.5~kpc for
 protons and 2~kpc for electrons \citep{Bloemen:1985,Bloemen:1986}. 
%0.5 for RSG and WR protons
%1 for ISM protons.
The spectral indices are approximated by constants throughout the
Galaxy and the normalization of the flux is taken from the review 
of \citet{Wiebel-Sooth:1998}.

The number of gamma-rays from a certain direction ($l$,$b$) with an energy
$E_\gamma$ per unit of time can then be calculated using
%\begin{eqnarray}
%\hspace{2cm} \frac{dn_\gamma(E_\gamma)}{dt dE_\gamma
%d\Omega}=\int{dL(l,b) \int{dE_k \int{d\epsilon}}} \nonumber
%\\
%\Phi_k(E_k,r,l,b) \sigma_i(E_k,E_\gamma) n_T(\epsilon,r,l,b)
%\end{eqnarray}
%
\begin{displaymath}
\hspace{-2cm} \frac{dn_\gamma(E_\gamma,l,b)}{dt dE_\gamma
d\Omega}=\int{dL(l,b) \int{dE_k }} \nonumber
\end{displaymath}\begin{displaymath}
\hspace{2cm} \Phi_k(E_k,r,l,b) \sigma_i(E_k,E_\gamma) n_T(r,l,b)
\end{displaymath}
where $\int{dL(l,b)}$ is the line-of-sight integral along the direction 
($l$,$b$),
$E_k$ and $\Phi_k(E_k,r,l,b)$ are energy and flux at position 
($r$,$l$,$b$) 
of the CR particles of type $k$, $\sigma_i(E_k,E_\gamma)$ is the 
cross section of the gamma-ray production process $i$, and  
$n_T(r,l,b)$ is the density of target particles of type $T$  
(either components of the ISM, photons of a radiation fields or particles 
of a SN-wind). In the case of IC it is also necessary to integrate over 
the energy distribution of the photon gas~$\epsilon$.
If we use the assumed independence of 
CR energy spectra from the position in our galaxy
\begin{displaymath}
\Phi_k(E_k,r,l,b) = \Phi_k^E(E_k) \cdot \Phi_k^s(r,l,b) ,
\end{displaymath}
the integral can get factorized into one integral over energy and one 
over the line-of-sight:
\begin{displaymath}
\frac{dn_\gamma(E_\gamma,l,b)}{dt dE_\gamma d\Omega}=I_E(E_\gamma) \cdot
I_L(b,l) ,
\end{displaymath}
with
\begin{equation}
\label{equ_IE}
I_E(E_\gamma)=\int{dE_k \Phi_k^E(E_k) \sigma_i(E_k,E_\gamma)}
\end{equation}
and
\begin{equation}
\label{equ_IL}
I_L(b,l)=\int{dL n_T(r,l,b) \Phi_k^s(r,l,b)} .
\end{equation}\\
The cross sections $\sigma_i$ for Bremsstrahlung and inverse Compton effect
have been numerically computed using the formulae given 
by \citet{Blumenthal:1970}. For nuclear reactions they have been 
simulated with DPMJET3 \citep{Roesler:2001}. \\
The independent solution for $I_E$ and $I_L$ are shown in Fig.~\ref{fig_production}
 and Fig.~\ref{fig_enhancement}.

%\begin{eqnarray*}
%flux_{\gamma} &=&  \sum_i \int \, d\,E_{i} \, \, \Phi^e_i(E_i) \sigma_{nucl} \, \int \, dL \, \, n(r,l,b) \, \,
%\Phi^s_i(r,l,b)   \\[2mm]
%& + & \int \, d\,E_{el} \, \, \Phi^{e_{el}}(E_{el}) \, \sigma_{brems}   \, \int \, dL  \,  n(r,l,b) \,\, \Phi^s_{el}(r,l,b)  \\[2mm]
%& + & \sum_{k} \, \int \, d\,E_{el} \, \, \Phi^{e_{el}}(E_{el}) \, \int \, d \epsilon \, \,\, \, u^{\epsilon}_k (\epsilon)\,\, \, \, 
%\sigma_{IC}\\[2mm]  
%&& \int \, d L  \, \, \,  u^s_k (r,l,b) \, \, \, \, \Phi^{e_{el}}(r,l,b) \, ,
%\end{eqnarray*}
%where $\sigma_{nucl}$ stands for the differential cross sections of nuclear production function via
%$\pi^0$ decay involving $i$ kind of nuclei and protons injected by the different sources. $\sigma_{brems}$ is 
%the differential cross sections of electron bremsstrahlung and 
%$\sigma_{IC}$ is the differential cross section of inverse Compton scattering. $\Phi^e_{i}(E_{i})$ is the energy spectrum
%and $\Phi^s_{i}(r,l,b) $ the spatial distribution of CR protons, while $\Phi^e_{el}(E_{el})$ is the energy spectrum
%and $\Phi^s_{el}(r,l,b) $ the spatial distribution of CR electrons. Finally $u^{s}_k(r,l,b)$ 
%and $u^{\epsilon}_k(\epsilon)$ are the spatial and energy density distributions of the different components of galactic radiation fields. 

\section{Spectral Analysis of the Gamma-Ray Emission}
%%The flux of diffuse gamma-ray emission from galactic latitude $b$ and longitude $l$ is obtained  \citep{Bertsch}
\begin{figure}[t]
\centering
\includegraphics[angle=270,width=8.0cm]{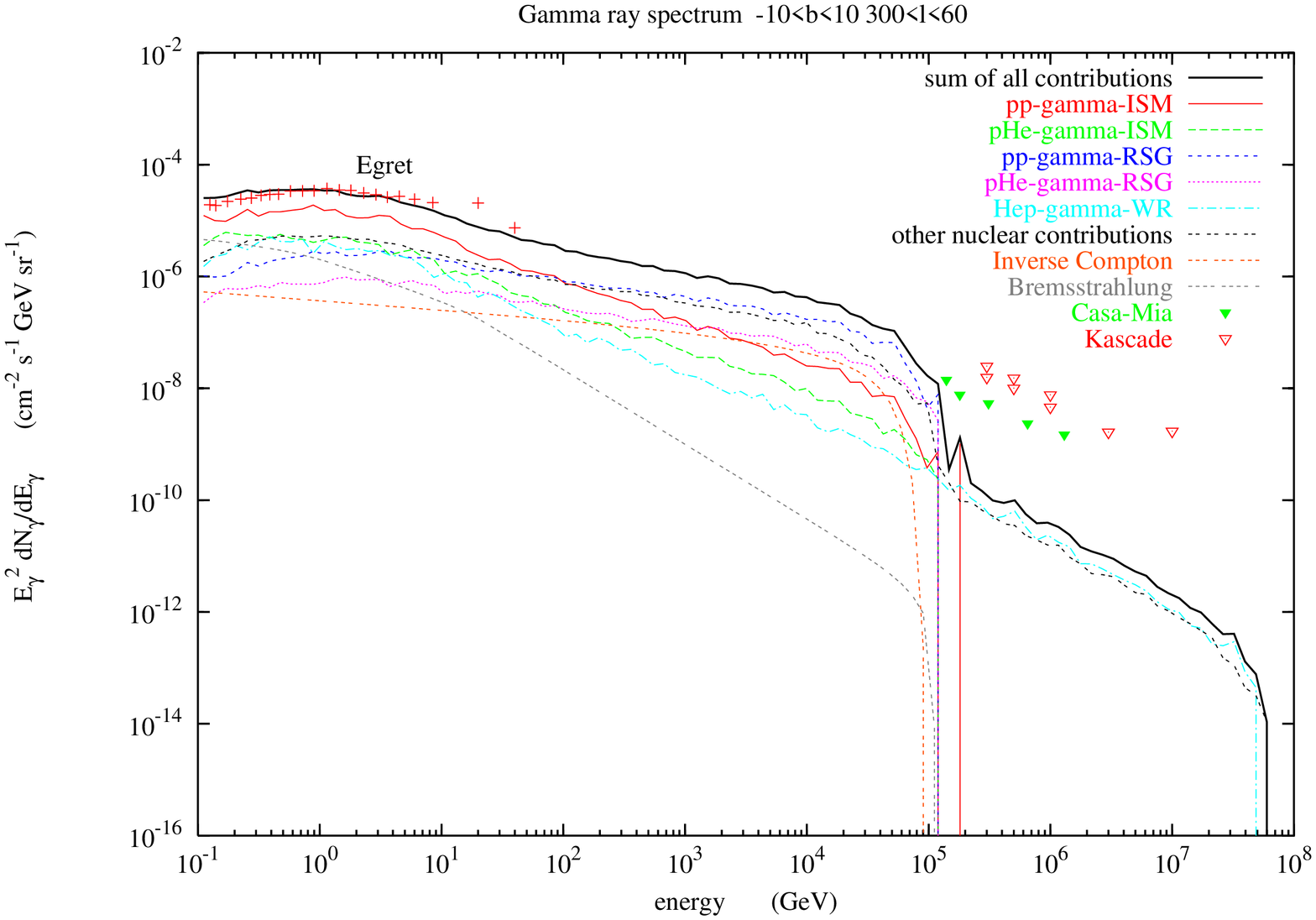} \\
\includegraphics[angle=270,width=8.0cm]{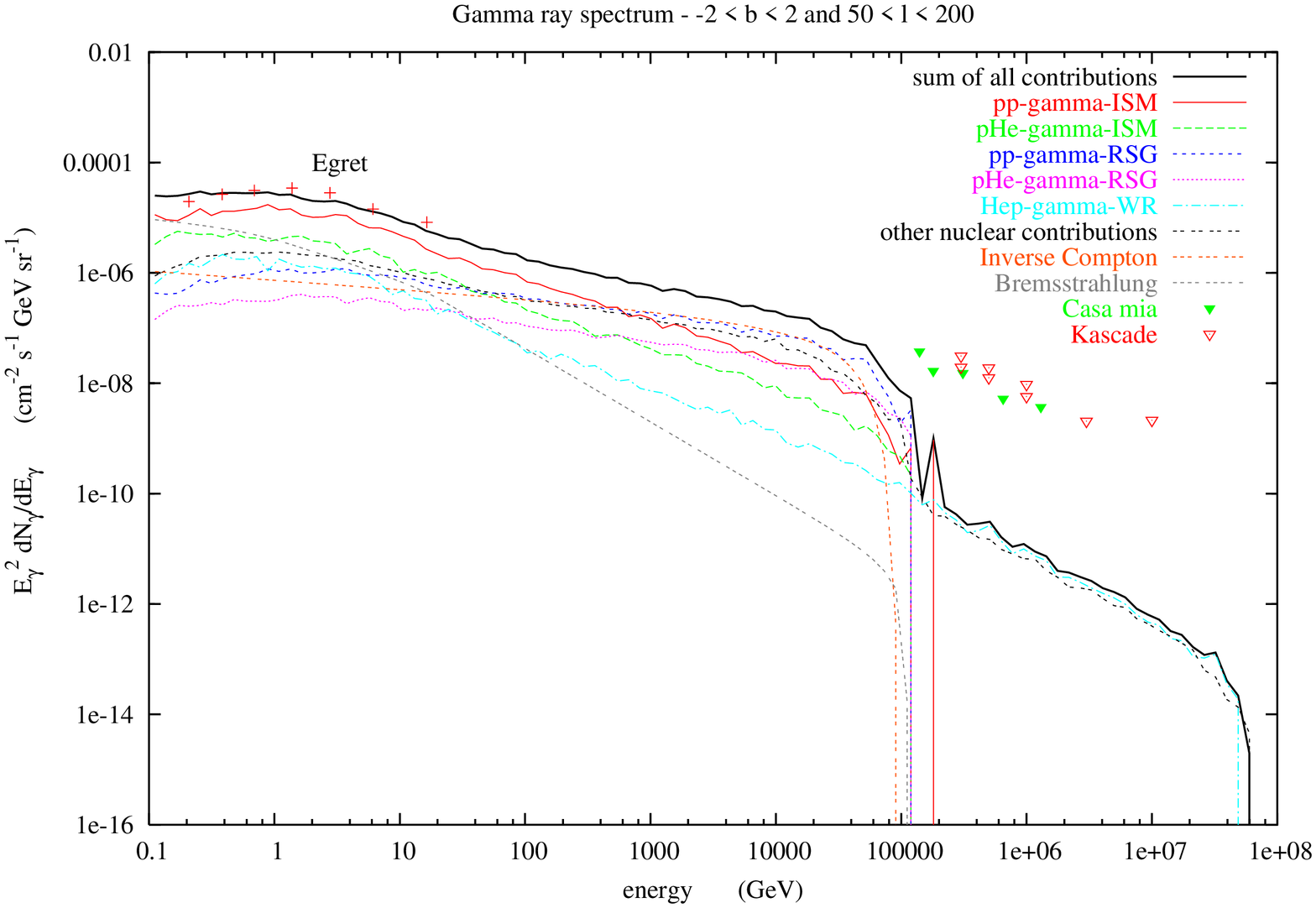}\\

\vspace{-10.5cm}
\hspace{0.5cm} \textbf{A}

\vspace{5.5cm}
\hspace{0.5cm} \textbf{B}

\vspace{4.4cm}

\caption{Gamma-ray energy spectra (complete model) for the directions of the galactic center (A)
and anti-center (B).}
\label{fig_spectra}
\end{figure}

The emission from the inner Galaxy and from the outer Galaxy, predicted by our model calculations, are shown 
in Fig.~\ref{fig_spectra} (A) and (B), respectively, where the predicted
spectra are compared to EGRET \citep{Hunter:1997}, Whipple \citep{LeBohec:2000},
 CASA-MIA \citep{Borione:1998} and KASCADE \citep{Schatz:2003} data. 
The contributions to the gamma-ray spectrum from the 
three kind of supernovae, from Bremsstrahlung and 
from inverse Compton scattering, as well as their sum, are explicitly 
indicated. 
The contribution to the diffuse gamma emission 
due to the collisions of CR protons with protons and helium 
in the interstellar medium explains the spectrum above 0.1~GeV
and below 1~GeV. Around 1~GeV the observed flux of diffuse galactic gamma-rays 
stems mainly from
interactions of CRs injected by both ISM and RSG supernovae.
Above 1~GeV it is produced essentially 
by interactions of protons injected by exploding RSG with the protons 
in the local strong and enriched winds.
Finally we remark that a change in the spectral index of the gamma-ray
emission is evident from Fig.~\ref{fig_spectra} (A) 
at photon energy above 1~GeV. Within our model the change in the 
gamma-ray spectral index arises naturally from the assumption that the two different regimes in 
the gamma-ray emission, corresponding
to gamma-rays emitted by different supernovae types.

\section{Directional Analysis of the Integrated Flux}
\begin{figure}
\centering
\includegraphics[width=8cm]{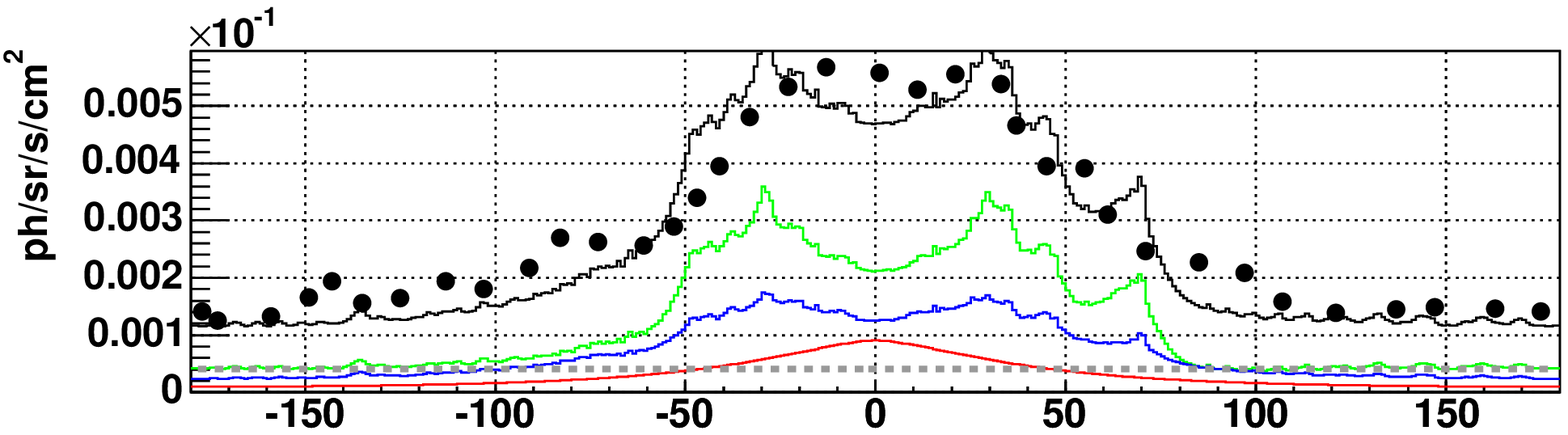}
\includegraphics[width=8cm]{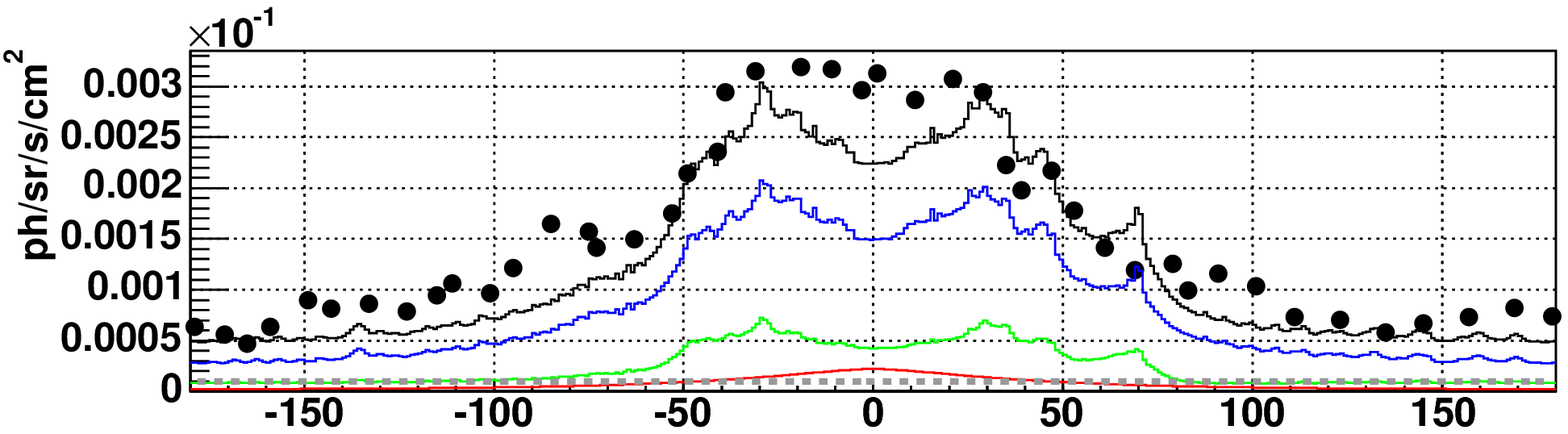}
\includegraphics[width=8cm]{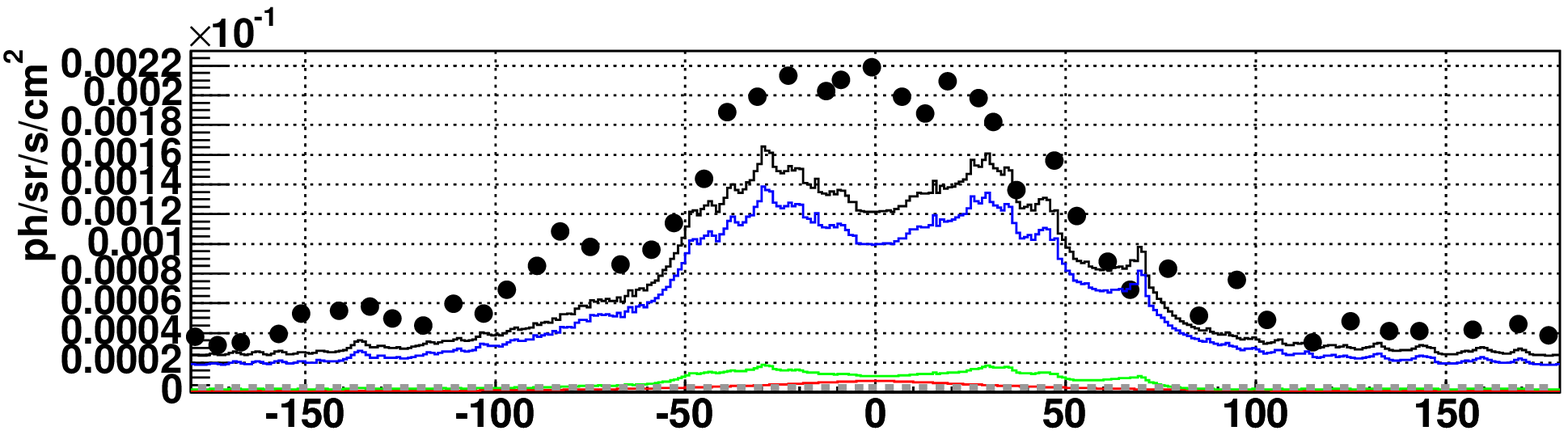}
\includegraphics[width=8cm]{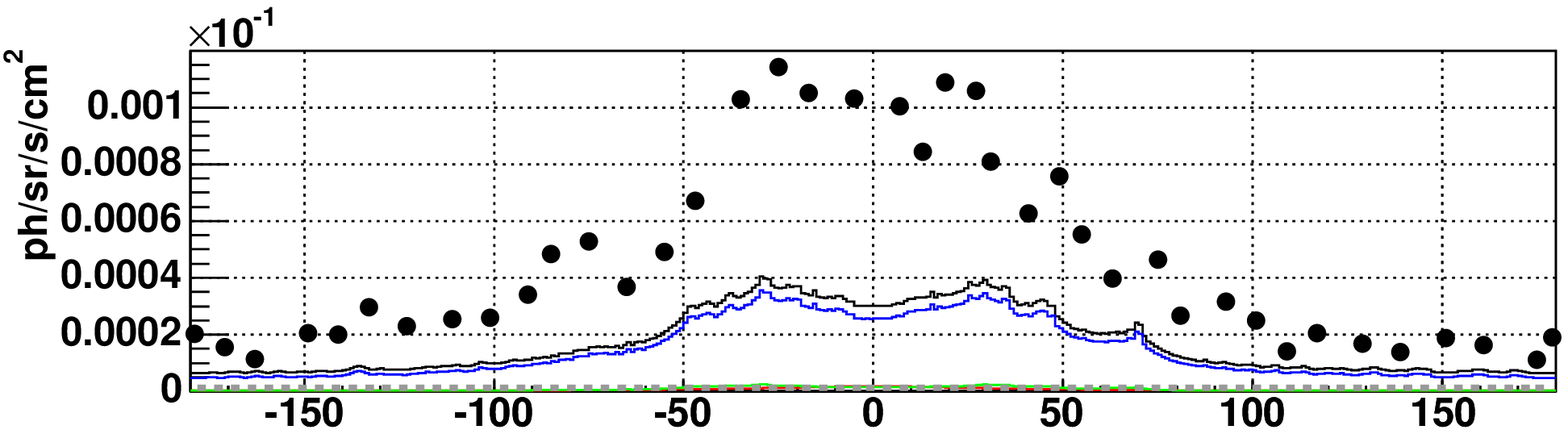}\\
\tiny

\vspace{-8.75cm}\hspace{3cm} {\bf a) 30MeV$<E_\gamma<$100MeV} \\
\vspace{1.95cm}\hspace{3cm} {\bf b) 100MeV$<E_\gamma<$300MeV} \\
\vspace{1.95cm}\hspace{3cm} {\bf c) 300MeV$<E_\gamma<$1000MeV} \\
\vspace{1.96cm}\hspace{3cm} {\bf d) 1000MeV$<E_\gamma<$30000MeV} \\
\vspace{2cm}

\normalsize
longitude

\caption{These plots only show the contribution to the gamma-ray flux
predicted by conventional models. The SN-wind 
component is \textbf{not} included.
Mean flux for $-2<b<2$. Different contributions to the longitudinal 
distributions: {\Red Inverse Compton}, 
{\Green Bremsstrahlung}, {\Blue Nuclear}, {\Gray Diffuse background}, {\Black
Sum}. The EGRET data is shown as black dots.}
\label{fig_long}
\end{figure}

The predicted longitude intensity distributions
of diffuse galactic gamma radiation are shown 
in Fig.~\ref{fig_long}. As can be seen in Fig.~\ref{fig_long} (a) the used 3D models for ISM, photons
and CR particles could be further optimized to reproduce the 
EGRET data even better in this energy range. More importantly, it is apparent
(Fig.~\ref{fig_long} (b)-(d)) that for
 higher energies the flux coming
from the Galactic center 
is getting more and more underestimated while the flux from the
anti-center stays relatively well reproduced. 
The discrepancy between the calculation and the EGRET data supports 
the picture of an additional contribution from CR
sources (SN) that are much more abundant in 
the inner Galaxy than in the outer.

\section{Discussion and Outlook}

The presented model is very promising in explaining the measured diffuse
 gamma-ray flux.
Also it seems to be capable to reproduce the C/B ratio as well as the
 antiproton flux \citep{Sina:2001}. 
It is quite a plausible scenario in which the CR interact mostly in the 
environment close to their sources. Some stars, like RSG or WR, provide
 sufficient 
material, ejected as powerful winds at the end of their lives, before they
 explode 
as supernovae. This wind material provides most of the grammage crossed 
by CR particles seen at Earth. We expect the data in the TeV range
 from MAGIC and HESS 
experiments to allow further insight or put constraints on the
 present model.
More detailed studies are planned to calculate the various predictions
of the model by \citet{Biermann:2001iu} and improve the 3D
of the interstellar matter, cosmic ray, and radiation field
distributions.
\\

\vspace{1cm}

\textit{The present work is being supported by AUGER theory and membership 
grant 05CU1ERA/3 through DESY/BMBF.}

\vspace{0.5cm}

%===============================================================================
% References
%===============================================================================

\bibliography{Proc}

\end{document}